\documentclass[%
prc,%
twocolumn,%
superscriptaddress,%
floatfix,%
twoside,
]{revtex4-2}

\usepackage[english]{babel}
\usepackage{graphicx}

\usepackage{amsmath}
\usepackage{amssymb}
\usepackage{amsfonts}
\usepackage{isotope}

\begin{document}
\title{Neutron versus proton scattering on exotic nuclei: the $^9$He example}

\author{M.S.~Khirk}
\email{mskhirk@jinr.ru}
\affiliation{Flerov Laboratory of Nuclear Reactions, JINR, 141980 Dubna,
Russia}

\author{L.V.~Grigorenko}
\affiliation{Flerov Laboratory of Nuclear Reactions, JINR,  141980 Dubna,
Russia}
\affiliation{National Research Nuclear University ``MEPhI'', 115409 Moscow,
Russia}
\affiliation{National Research Centre ``Kurchatov Institute'', Kurchatov sq.\ 1,
123182 Moscow, Russia}

\author{D.E.~Lanskoy}
\affiliation{Faculty of Physics, Lomonosov Moscow State University, Leninskie 
Gory,
Moscow 119991, Russia}

\author{P.G.~Sharov}
\affiliation{Flerov Laboratory of Nuclear Reactions, JINR,  141980 Dubna,
Russia}
\affiliation{Institute of Physics in Opava, Silesian University in Opava, 74601 
Opava,
Czech Republic}


\begin{abstract}
Neutron scattering on exotic nuclides is a class of processes which cannot be
studied directly now and in any observable future. Resonance proton scattering
of exotic nuclide on a thick target in inverse kinematics can be used to infer
the properties of the low-energy neutron scattering of this nuclide assuming the
isobaric symmetry. However, the results of such resonance proton scattering
reactions are so far analyzed in theoretical approaches (optical, R-matrix
models), which are missing important aspects of isospin dynamics, isospin
violation in continuum and threshold dynamics. The isospin
conserving coupled-channel model (ICM) is proposed, which provides a more
reliable basis for understanding of such experimental studies. Qualitatively
different phase shifts for the $^{8}$He+$p$ $T=5/2$ and $T=3/2$ resonances are
predicted by ICM with quite unusual profile for the $T=5/2$ states. Alternative
interpretation of the existing $^{8}$He+$p$ data is proposed. The observable
properties of the $T=5/2$ resonances may be strongly affected by the
isobaric-partner $T=3/2$ states. Crucial importance of studies of the
neutron-emission channel for disentangling this possible influence is
demonstrated.
\end{abstract}


\maketitle

\textit{Introduction.}
%
Resonance proton scattering (RPS) on a thick target in inverse kinematics is an
elegant and powerful experimental method \cite{Goldberg:1998}. It was found
especially efficient for studies with exotic radioactive beams of low quality
and intensity, because very thick targets can be used: the low-energy elastic
scattering excitation functions in a broad energy range and corresponding angular distributions are
obtained simultaneously with a fixed-energy incoming beam. The application of
the method is very natural  for the proton continuum of \emph{proton-rich}
exotic nuclei, where the interpretation of elastic scattering results is
straightforward and unique
\cite{Goldberg:1998,Lee:2007,Mountford:2012,DeGrancey:2016,Girard:2022}.

Another application of the method presumes getting information on neutron
scattering of some \emph{neutron-rich} exotic nuclide $^AZ$+$n$ by studies of
the ``isobaric partner reaction'' of proton scattering $^AZ$+$p$. For each
spin-parity $J^{\pi}$ the observed isobaric-analog state (IAS) in
$^AZ$+$p$ channel with $\{T=T_{\max},T_3=T_{\max}-1\}$ is described by some
theoretical model. By switching off the Coulomb interaction in the
$^AZ$+$p$ channel the properties of the $^AZ$+$n$ continuum with
$\{T=T_{\max},T_3=T_{\max}\}$ are deduced
\cite{Rogachev:2003a,Rogachev:2004,DeOliveira:2009,Hunt:2020,Uberseder:2016,%
Hunt:2023}.

The results of $^{8}$He+$p$ studies aiming $^{9}$He properties were analyzed in
models (optical potential, R-matrix) \cite{Rogachev:2003a,Uberseder:2016},
which have some room for phenomenological treatment of channel coupling, but are
missing specific aspects of isospin dynamics connected with isospin
conservation. The following issues are shown in this work to be important
specifically for $^{8}$He+$p$ studies and may be important for application of
the RPS method in general:

\begin{figure}
\begin{center}
\includegraphics[width=0.49\textwidth]{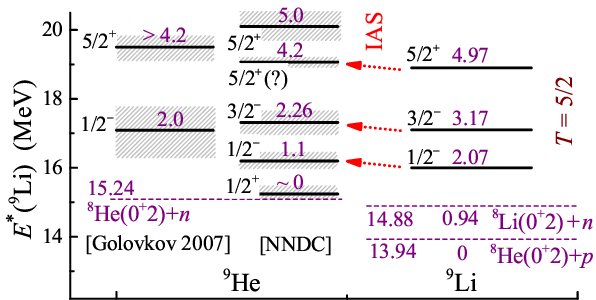}
\end{center}
\caption{Relevant levels in $^{9}$He and $^{9}$Li systems.
The data of \cite{Golovkov:2007} is shown separately since it provides quite 
different from NNDC vision of the low-lying negative-parity states.}
\label{fig:levels}
\end{figure}


\noindent (i) The continuum states in $^AZ$+$p$ channel with
$\{T=T_{\max},T_3=T_{\max}-1\}$ are typically ``globally'' (in the whole radial
space)
isospin-mixed. This is always true for $^AZ$ systems with nonzero total isospin.
For such cluster-continuum configurations the good quantum number of
isospin is recovered ``locally'' (in some radially-limited ``nuclear
interior''). To make this recovery possible the mixing of the $^AZ$+$p$ channel
with  $^A(Z+1,\text{IAS})$+$n$ channel is needed. In our $^{9}$He case these are
$^{8}$He($0^+2)$+$p$ and $^{8}\text{Li}^*(0^+2)$+$n$ channels. The aspect of
this mixing governed by the isospin conservation can be reliably modelled.

\begin{figure*}
\begin{center}
\includegraphics[width=0.99\textwidth]{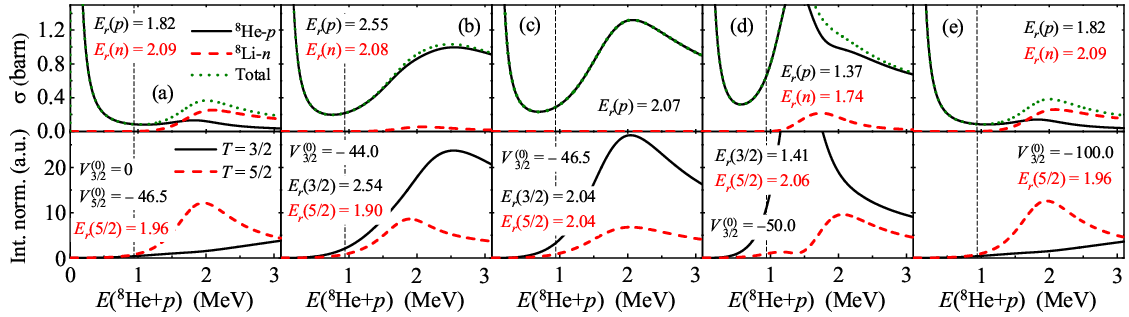}
\end{center}
\caption{Upper panels show cross sections for
the elastic $^{8}$He+$p$ and inelastic $^8\text{Li}^*(0^{+}2)$+$n$ channels.
Lower panels show the isospin content of the continuum states in terms of
internal normalizations up to 6 fm. Columns (a)--(e) correspond to
different cases of $V_{3/2}$ interactions. Value  $E_r(i)$ gives the visible peak position in the ``$\textit{i}$''-th channel.
Resonant energy of the $1/2^-$ state in $^9$He is fixed at  $E_r = 1.1$
MeV ($V_{5/2}^{(0)}=-46.5$ MeV, $r_0=2.3$ fm). The $^8\text{Li}^*(0^{+}2)$+$n$
threshold is shown by vertical dashed lines. All the values of energies in the 
legends are given in MeV.}
\label{fig:mult}
\end{figure*}

\noindent (ii) The pure isospin concept is perfect for zero-width discrete
states and quite precise for small-width continuum states. The states in the
dripline neutron-rich systems, e.g.\ $^{9}$He=$^8$He+$n$, are expected to be
quite broad. This means that the corresponding states in the  $^{8}$He+$p$
continuum should be even broader. There is certain isospin-symmetry-violation
aspect of nuclear dynamics, which is connected just with large width of the
considered states: broad states with definite isospin are stronger coupled to
both $^{8}\text{He}(0^+2)$+$p$ and $^{8}\text{Li}^*(0^+2)$+$n$ continuums. Each
of these continuums do not have definite isospin. So, the isospin mixing
stemming from the continuum couplings becomes the more important the broader are
considered states.

\noindent (iii) Mixing of the $^AZ$+$p$ and $^A(Z+1,\text{IAS})$+$n$
channels presume that together with the $T=T_{\max}$ IAS state, an ``isobaric
partner'' state with $T=T_{\max}-1$ may be existing nearby in energy. In the
spectra of $^{4}$He and $^{8}$Be such isospin doublets are known and well-
studied, e.g.\ \cite{Grigorenko:2000,Grigorenko:2000a} and Refs.\ therein. In
the case of insufficient energy separation the ``isobaric partner'' state should
significantly affect the observable properties of the $^AZ$+$p$ IAS state. We
show in this work that for states broader than $\Gamma \gtrsim 0.3$ MeV any
realistic energy gap between them is just insufficient to make their
interference negligible. This is making
interpretation of the proton-scattering data on broader states in terms of
isolated resonances unreliable.

\noindent (iv) In the light nuclei the thresholds of the $^AZ$+$p$ and
$^A(Z+1,\text{IAS})$+$n$ channels are typically located just within $\sim$ 1 MeV of
energy. For broad states complicated threshold dynamics may take place around
these thresholds. For example, dynamical studies of these phenomena in the
coupled-channel model are absolutely essential for understanding of $l=0$
states: they are represented by \emph{resonances} in the $^AZ$+$p$ continuum,
but by \emph{virtual states} $^AZ$+$n$ and  $^A(Z+1,\text{IAS})$+$n$ channels,
which make the situation extremely complicated for interpretation.


The above aspects of isospin dynamics can be accounted in a consistent way in a
relatively simple coupled-channel model. Such an isospin-conserving model
was developed and successfully applied for studies of the isospin mixing in
$^{4}$He and $^{8}$Be systems in Refs.\ \cite{Grigorenko:2000,Grigorenko:2000a}.
We demonstrate below that basing on such a model the experimental data may be
interpreted very differently and, moreover, the reliable interpretation is
possible only if the $^AZ$+$p$ scattering data is augmented with the
neutron-emission $^A(Z+1,\text{IAS})$+$n$ channel data.

\begin{figure*}
\begin{center}
\includegraphics[width=0.98\textwidth]{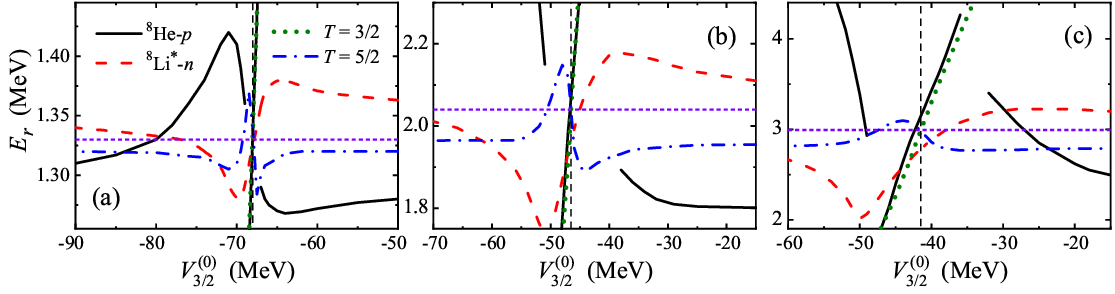}
\end{center}
\caption{Trajectories of the peak values for elastic $^{8}$He+$p$, inelastic
$^8\text{Li}^*(0^{+}2)$+$n$ channels and isospin content of the continuum
states. Panels (a), (b), and (c) correspond to the resonant energy  $E_r$ of
the $1/2^-$ state in $^{9}$He equal to 0.4, 1.1 and 2.1 MeV respectively, 
$V^{(0)}_{5/2}=\{-68.0, -46.5, -41.5\}$ MeV. The 
vertical dashed lines indicate the degeneracy situation $V_{3/2}\equiv V_{5/2}$; 
the horizontal dotted lines show the resonant peak position (from the same threshold) in elastic 
$^8\text{Li}^*(0^{+}2)$+$n$ channel for this case.
}
\label{fig:traject}
\end{figure*}

\textit{Situation with understanding of the $^{9}$He spectrum}
%
is one of motivations to use the $^{9}$He system as example. This situation is
quite controversial, see nice summary of the data in \cite{Kalanee:2013}, some
more recent data in \cite{Votaw:2020}, and also Fig.\ \ref{fig:levels}.

There is some common agreement about positioning of the $5/2^+$ state but with
quite a large energy uncertainty $E_r \sim 3.4-5.2$ MeV. Also there is some 
common
agreement about positioning of the $1/2^-$ state $E_r \sim 1.1-1.3$ MeV.
However, the only work, which provides spin-parity identification, positions
$1/2^-$ quite differently with $E_r = 2 \pm0.2$ MeV  \cite{Golovkov:2007}. The
NNDC prescription, see also \cite{Tilley:2004}, for the $T=5/2$ states
located high in the $^{9}$Li continuum comes only from the $^{8}$He+$p$ data of
Ref.\ \cite{Rogachev:2003a}. Also the possibility of the low-lying $3/2^-$ state
in $^{9}$He was inferred basing on these data.

The possibility of the low-lying $1/2^+$ has been considered many times in the
literature since some evidence for a large negative 
scattering length $a_{\text{s}}<-10$ fm was found in Ref. \cite{Chen:2001}. 
Nevertheless, the situation remains uncertain. On the one hand, a low-lying structure, corresponding to
$a_{\text{s}} \approx -12 \pm 3$ fm was reported in \cite{Kalanee:2013}. On the
other hand, no ``strong'' virtual state in $^{9}$He was found in Refs.\
\cite{Johansson:2010,Falou:2011} providing the scattering lengths $a_{\text{s}}
\approx -3$ fm and $a_{\text{s}} \gtrsim -3$ fm, correspondingly. Also, in Ref.\
\cite{Golovkov:2009} the large negative scattering length, e.g.\ as large as
$a_{\text{s}} \approx - 20 $ fm is, in principle, not completely excluded, but
is highly unfavorable. In the analysis \cite{Votaw:2020} the
$^{8}$He+$n$ final state interaction allows both modest  $a_{\text{s}} \sim -2$ 
fm and relatively strong $a_{\text{s}} \sim -10$ fm depending on assumptions. 
Theoretical analysis of \cite{Sharov:2014} based simultaneously on the $^{9}$He 
and $^{10}$He data provides the limitation $a_{\text{s}} \gtrsim 1$ fm.

Vision of the $^{9}$He spectrum which is strongly different from anything listed
above is proposed based on the $^{8}$He+$p$ data in Ref.\ \cite{Uberseder:2016}:
the $1/2^+$ resonance at $E_r \sim 3$ MeV and simultaneously no room for the
low-lying $p$-wave resonance with $E_r \lesssim 2.0-2.5$ MeV.

In these situation of the long-term confusion concerning the $^{9}$He spectrum
the additional $^{8}$He+$p$ data may provide crucial missing information.
However, it could be done only if the issues of the isospin mixing are
theoretically reliably resolved.

\begin{figure}
\begin{center}
\includegraphics[width=0.485\textwidth]{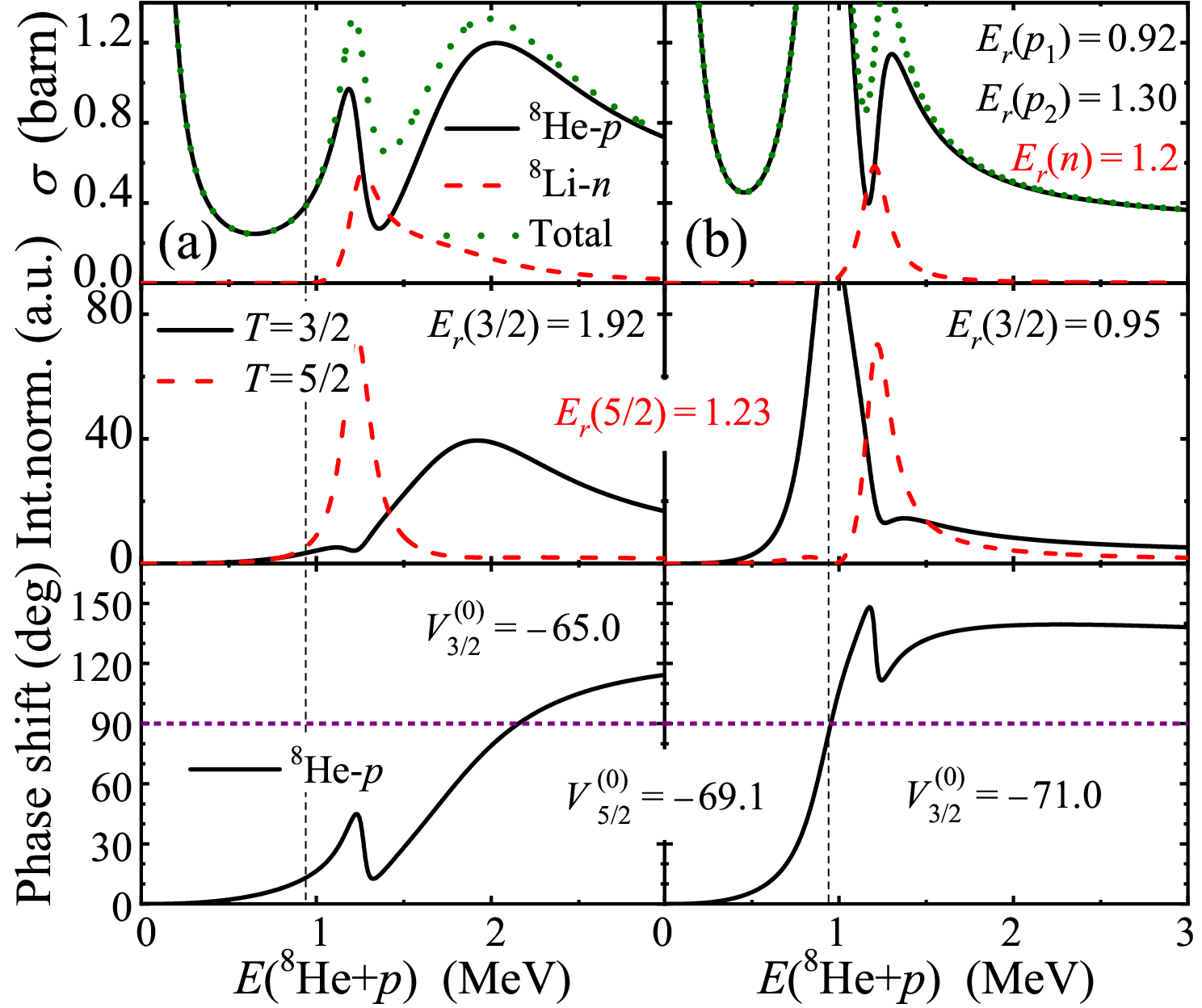}
\end{center}
\caption{Cross sections, isospin populations and phase shifts in the
$^{8}$He+$p$ channel. Case of $E_r(1/2^-)=0.3$ MeV in $^{9}$He
($V^{(0)}_{5/2}=-69.1$ MeV, $r_0=2.0$ fm) and the $V_{3/2}$
interaction providing a small split between $T=3/2$ and  $T=5/2$ states. Column
(a) corresponds to $E_r(5/2)< E_r(3/2)$, and (b) corresponds to $E_r(5/2) >
E_r(3/2)$. All the values of energies in the legends are given in MeV.}
\label{fig:phases}
\end{figure}

\textit{Isospin-conserving model for the $^8$He-$p$ reaction.}
%
Let us rewrite model \cite{Grigorenko:2000} for the case of the $^8$He+$p$
resonance scattering reaction. For the $^8$He(0$^{+}2)$+$p$ and
$^8\text{Li}^*(0^{+}2)$+$n$ channels the assumption of isospin symmetry leads to
interaction between clusters which can be represented as a sum of terms with
definite isospin:
\[
\hat{V}=V_{3/2}(r)\,\hat{P}_{3/2}+V_{5/2}(r)\,\hat{P}_{5/2} \, ,
\]
where $\hat{P}_{T}$ are projection operators on the states with definite total
isospin $T$. The cluster WFs with definite asymptotic conditions are connected
to WFs with definite isospin $\left|  T, T_3 \right\rangle$ as
\begin{equation}
\left\{
\begin{array}
[c]{l}%
\Psi_{^8\text{He-}p} =\frac{1}{\sqrt{5}}\left|  5/2,3/2 \right \rangle +\frac
{2}{\sqrt{5}} \left|  3/2,3/2 \right \rangle  \\
\Psi_{^8\text{Li}^*\text{-}n} =\frac{2}{\sqrt{5}}\left|  5/2,3/2\right \rangle
-\frac
{1}{\sqrt{5}}\left|  3/2,3/2 \right \rangle
\end{array}
\right.\, .
\label{eq:chan-via-isosp}
\end{equation}
From this decomposition it is clear that coupling of the scattering $^{8}$He+$p$
channel to the $T=5/2$ is relatively weak, which appears to be very important, see Fig.\
\ref{fig:phases} and discussion around.

By diagonalizing the Schr\"odinger equation we get a system of coupled equations
\begin{equation}
\begin{split}
&\hspace{-30pt}\left[  \hat{T}-(E-\Delta E)+(1/5) \left( V_{3/2}+4V_{5/2} \right) \right]
\Psi_{^8\text{Li}^*\text{-}n} \\
&\hspace{40pt}+(2/5) \left( V_{5/2}-V_{3/2} \right) \Psi_{^8\text{He-}p} = 0  \,,
\\
&\hspace{-30pt}\left[  \hat{T}-E+V_{\text{coul}}+(1/5) \left(4V_{3/2}+V_{5/2} \right) \right]
\Psi_{^8\text{He-}p}\\
&\hspace{40pt}+ (2/5) \left(  V_{5/2}-V_{3/2} \right) \Psi_{^8\text{Li}^*\text{-}n} = 0
\,,
\label{eq:schr}
\end{split}
\end{equation}
where $E$ is the threshold energy in the $^{8}$He+$p$ channel and
$\Delta E = 0.941$ MeV is threshold shift of the $^8\text{Li}^*(0^{+}2)$+$n$ 
channel. The interaction $V_{5/2}$ is what we are intended to determine to fix 
the spectrum of $^{9}$He. However, this can be done only if the $V_{3/2}$ 
interaction is fixed somehow.

The potentials with the Gaussian formfactors
\[
V_T(r)=V_T^{(0)} \, \exp[-(r/r_0)^2]\,,
\]
are used in the calculations. The Coulomb interaction of the homogeneously
charged sphere is used with radius 2.5 fm, consistent with the charge radius of
$^{8}$He 1.956(16) fm \cite{Mueller:2007,Krauth:2021}. The potential radius $r_0
\sim 2.0-2.4$ fm is fine-tuned by condition that the resonant states in both the
$^{8}\text{He}^*(0^{+}2)$+$p$ and $^8\text{Li}^*(0^{+}2)$+$n$ decoupled channels
(case $V_{3/2} \equiv V_{5/2}$) have the same energy near the $^{8}$He+$p$
threshold. This guarantees that the radial properties of the nucleon
orbitals in the $^{8}$He$^*(0^{+}$2)+$p$, $^8$Li$^*$+$n$ channels and inside the
$^{8}$He, $^8\text{Li}^*$ systems are consistent with the threshold energy
shift. This is a reasonable isobaric-symmetry requirement.


\textit{General features of the model and guidelines for experiment.}
%
The evolution of the $^{8}$He+$p$ scattering with variation of the $V_{3/2}$
interaction is illustrated in Fig.\ \ref{fig:mult} for the $1/2^-$ continuum.
Cross sections for the elastic $^{8}$He+$p$ and inelastic
$^8\text{Li}^*(0^{+}2)$+$n$ channels are evidently observables and isospin
contents of the continuum states could also be related to observables (e.g.\
strength functions for isospin-specific reactions).

It can be seen from Eq.\ (\ref{eq:schr}), that reduction to the single channel
formulation takes place at $V_{3/2}\equiv V_{5/2}$, Fig.\ \ref{fig:mult} (c). It
is somehow paradoxical, but such highest ``isospin symmetry'' leads to isospin
degeneracy and mixing. For relatively large differences $\vert V_{3/2} - V_{5/2}
\vert$ the isospin symmetry is recovered in our model dynamically with a good
precision, see Figs.\ \ref{fig:mult} (a), (e). In between of these limiting situations the
isospin mixing effects are severe and complicated, see Figs.\ \ref{fig:mult}
(b), (d), and dynamical calculations are absolutely necessary.

Trajectories in the $\{V_{3/2},E\}$ plane for the peaks of different
observables, are summarized in Fig.\ \ref{fig:traject}. This is done for three
cases of the $1/2^-$ state in the $^{9}$He: (a) $E_r= 0.4$, $\Gamma\approx 0.2$ 
MeV --- test case of narrow resonances, (b) $E_r= 1.1$, $\Gamma\approx 0.8$ MeV 
--- as in NNDC, see Fig.\ \ref{fig:levels}, and (c) $E_r = 2.1$, $\Gamma\approx 
1.9$ MeV --- as in Ref.\ \cite{Golovkov:2007}. The following should be noted 
here:

\noindent (i) The scale of energy variation of different peaks with variation of $V_{3/2}$ interaction is
of the order of $\Gamma/2$, which is quite expectable. This is \emph{large}
effect for broad states (expected in the $^{9}$He case), which cannot be just
disregarded.

\noindent (ii) The deviation of different peaks in Fig.\ \ref{fig:traject} from 
the expected in simple approximation $T=5/2$ isobaric peak position (dotted 
lines in Fig.\ \ref{fig:traject}: $^8$Li$^*(0^{+}$2)+$n$ decoupled case) 
is practically never negligible: the realistic 
separation of, say, $2-5$ MeV from the  $T=3/2$ state is not sufficient to 
completely get rid of its effect on the $T=5/2$ state position observed in 
different channels.

\noindent (iii) The peak values demonstrate ``asynchronous'' behavior with
variation of $V_{3/2}$. It looks like the $T=3/2$ state ``repels'' the
$^{8}$He+$p$ continuum and ``attracts'' the $^8$Li$^*(0^{+}$2)+$n$ continuum.
This ``asynchronicity'' means that \emph{one single type of the data} is not sufficient to
understand what is the actual $T=5/2$ position, which actually stays constant in
each calculation of Fig.\ \ref{fig:traject}. Only the studies of the
$^{8}$He+$p$ and $^8\text{Li}^*(0^{+}2)$+$n$ channels simultaneously may provide
sufficient evidence to fix the $T=5/2$ properties if one pretends better than $\Gamma/2$
precision.


\textit{The phase shift issue.}
%
Important feature of the phase shifts obtained in the ICM are illustrated in Fig.\ \ref{fig:phases}
for the case of quite narrow $1/2^-$ resonances, which is easiest for
perception.
The $T=5/2$ resonance position is fixed at $E_r(5/2)=1.23$ MeV. The use of somewhat 
deeper or somewhat weaker  $V^{(0)}_{3/2}$ interaction provides the $T=3/2$ resonance 
position nearby.
One may see that whatever state is higher in energy --- the $T=3/2$
in Fig.\ \ref{fig:phases} (a) or the $T=5/2$ in Fig.\ \ref{fig:phases} (b) ---
the $T=3/2$ state demonstrates the ``classical'' resonant behavior, with phase
shift passing $\pi/2$, while at the $T=5/2$ resonance energy only a relatively
small ``wiggle'' in the phase shift is observed. This wiggle is a typical
interference
pattern for weakly-coupled resonance amplitude with exactly $\pi/2$ ``initial''
relative phase. For the broader states the $T=5/2$ wiggles become difficult to
identify, while the resonant behavior of phase at the $T=3/2$ state energy
persists.

The single-channel optical model phase shifts in Ref. \cite{Rogachev:2003a} and
somehow combined optical model and R-matrix  phase shifts in Ref.
\cite{Uberseder:2016} were used for analysis of the $^{8}$He-$p$ data. These
approaches rely on a standard resonance identification procedure: the large
variation of phase shift passing $\pi/2$ (or close to that).  It can be seen in
Fig.\ \ref{fig:phases} that in the model with
appropriate isospin treatment such a behavior is associated only with the
$T=3/2$ states. We have to conclude here, that the resonant properties were
likely misinterpreted and the states declared as $T=5/2$ due to the phase shifts should be in reality
the $T=3/2$ states. This concerns the listed in NNDC 16.0, 17.1, 18.9 MeV states
of $^{9}$Li, all interpreted as $T=5/2$, and to 2.26, 4.2 MeV states of
$^{9}$He, inferred by isobaric symmetry, see Fig.\ \ref{fig:levels}.

Such phase shift behavior in the ICM is generic. It is always present on some level for
proton channels in situation $T>1$. So, we may foresee the importance of such
model studies for analysis of the RPC data on other neutron-rich systems as
well.

\begin{figure}
\begin{center}
\includegraphics[width=0.40\textwidth]{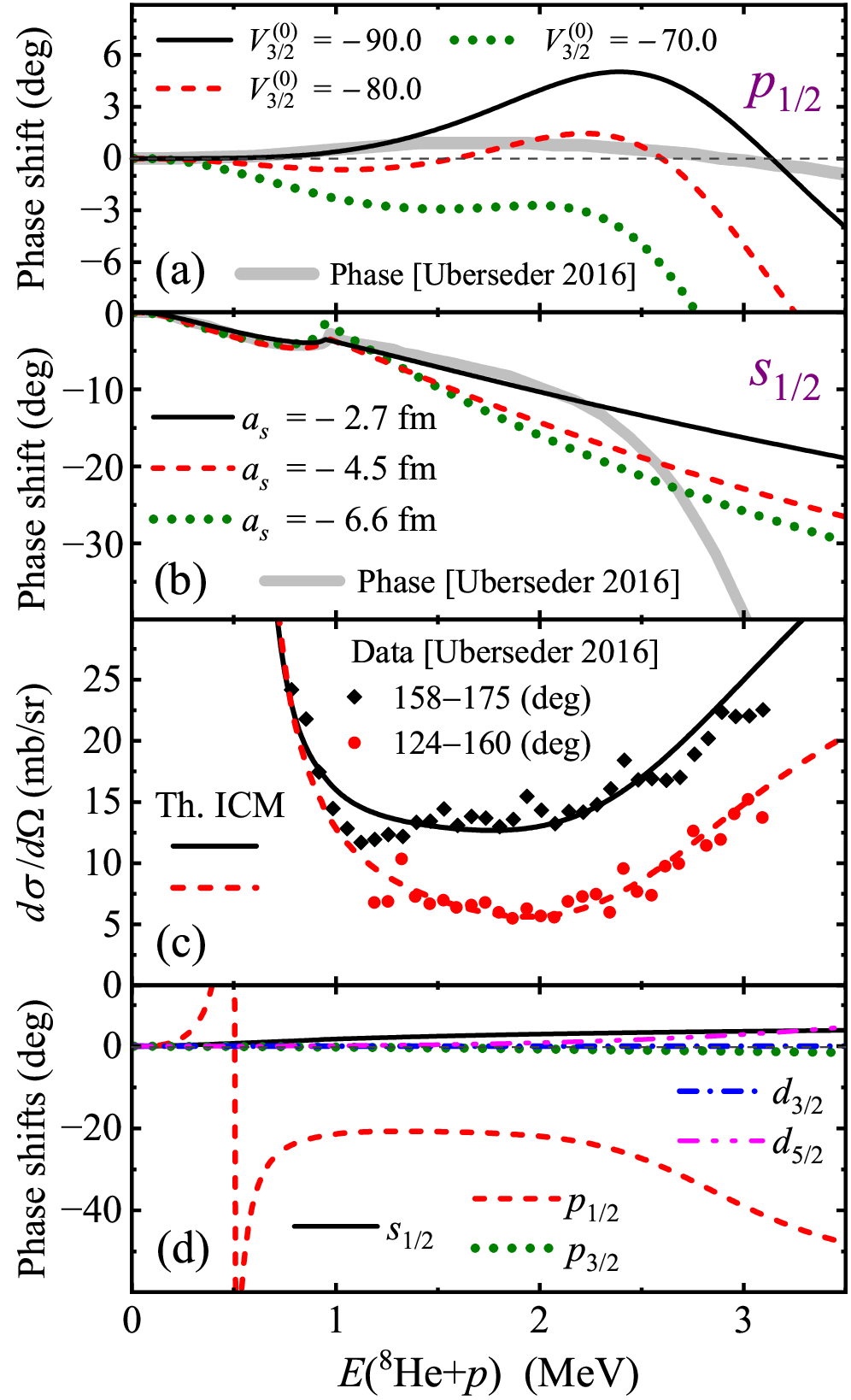}
\end{center}
\caption{Panel (a) shows the ICM $p_{1/2}$ phase shifts based on the 
$T=5/2$ resonance in $^{9}$He $E_r(1/2^-) = 2.1$ MeV 
($V^{(0)}_{5/2} = -41.5$ MeV), compared with phase used in analysis
of \cite{Uberseder:2016}. Panel (b) shows the ICM $s_{1/2}$ phase shifts
possessing the Wigner cusp, analogous to that used in analysis of
\cite{Uberseder:2016}. They are repulsive overall
(negative phase shifts, $V^{(0)}_{3/2}=\{-52, -48, -47\}$ MeV), while attractive interactions 
(negative scattering lengths for the $T = 5/2$ interactions $V^{(0)}_{5/2}=\{-6.0, -7.5, -8.5\}$ MeV are indicated in the panel). 
Panel (c) shows a variant of the ICM analysis of data \cite{Uberseder:2016} based on the
phases shown in panel (d). The data and fit in (c) for the $124-160$ degrees
center-of-mass angular range are shown with $-5$ mb/sr offset to simplify perception.
All the values of energies in the legends are given in MeV.}
\label{fig:exp-fit}
\end{figure}


\textit{Discussion of the $^{8}$He+$p$ data interpretation in
\cite{Uberseder:2016}.}
%
%
In the analysis of that work the very close to zero phase shifts were deduced
for the $p_{1/2}$, $p_{3/2}$, $d_{3/2}$ and $d_{5/2}$ configurations. The
$d_{5/2}$ resonance in $^{9}$He is expected to be quite high in energy and its
manifestation in the energy window $E(^8$He+$p)<3.2$ MeV accessible in
experiment \cite{Uberseder:2016} may be quite small. However, it is amusing that
there is no indication of the $p_{1/2}$ resonance at all. It is shown in Fig.\
\ref{fig:exp-fit} (a) that in the ICM the very small $p_{1/2}$ phases in the
whole  $E(^8$He+$p)<3$ MeV energy range may be obtained despite the presence in this range of
the $p_{1/2}$ resonance with $T=5/2$, corresponding to the $1/2^-$ resonance in
$^{9}$He at $E_r \sim 2.1$ MeV. Technically, such a small phase shift in a broad energy 
range is obtained by combining the "wiggle" feature associated with $T=5/2$ 
(rising phase shift trend, see Fig.\ \ref{fig:phases} and related text) with repulsion in the 
$T=3/2$ channel (negative and decreasing phase shift in contrast with illustrations of Fig.\ \ref{fig:phases}).
This situation is consistent with the $^{9}$He
data of \cite{Golovkov:2009}, providing $E_r(1/2^-) = 2.0 \pm 0.2$ MeV.

The structured character of the $^{8}$He+$p$ cross section in the energy range
$0.8<E(^8$He+$p)<3.2$ MeV is mainly related in the interpretation of Ref.\ \cite{Uberseder:2016} to a
curious behavior of the  $s_{1/2}$ phase shift. This demonstrates repulsion near
the $^{8}$He+$p$ threshold, Wigner cusp at the $^{8}$Li$^*(0^{+}$2)+$n$ threshold and
$1/2^+$ resonance with $E_r \gtrsim 3.2$ MeV, see Fig.\ \ref{fig:exp-fit} (b).
The latter is interpreted in \cite{Uberseder:2016} as corresponding to $1/2^+$
resonance in $^{9}$He at $E_r \sim 3$  MeV. It should be noted that existence of
the $s_{1/2}$ resonance in Ref.\ \cite{Uberseder:2016} is directly related to
the cusp properties. However, it is
shown in Fig.\ \ref{fig:exp-fit} (b)  that in the ICM the analogous cusp
behavior can be obtained for variety of situations without any need for $1/2^+$
resonance, including the situation of quite weak attraction in $1/2^+$ channel,
reasonably consistent with the theoretical limitations $a_s \gtrsim 0$ fm
deduced in Ref.\ \cite{Sharov:2014}.

Generally, we can get in ICM very good fits to the data of Ref.\
\cite{Uberseder:2016}, see, for example, Fig.\ \ref{fig:exp-fit} (c).
This fit is based on the phases Fig.\ \ref{fig:exp-fit} (d), obtained with
$T=5/2$ interaction, which is consistent with the $^{9}$He data of
\cite{Golovkov:2009}: weak attraction is $1/2^+$ channel, $1/2^-$ resonance $E_r
= 2.3$ MeV, $5/2^+$ resonance $E_r = 4.7$ MeV. The parameters of the interactions 
are $V^{(0)}_{3/2}=\{0.0, -55.4, 5.0, 0.0, -60.0\}$ MeV and $V^{(0)}_{5/2}=\{-4.0, -40.5, 5.0, 0.0, -96.5\}$ MeV
for the $\{s_{1/2}, p_{1/2}, p_{3/2}, d_{3/2}, d_{5/2}\}$ quantum number sets.
We conclude here that the unconventional proposal of \cite{Uberseder:2016} --- (i) the low-lying $1/2^+$
resonance and (ii) no room for the low-lying $1/2^-$ resonance at all --- is actually based on the limited character of
the model used in \cite{Uberseder:2016}, which neglects complexity of the isospin
mixing dynamics.

The provided in Fig.\ \ref{fig:exp-fit} (d) analysis is not unique. Other fits
of analogous quality are possible including the cusp-inducing weak repulsion,
similar to cases of Fig.\ \ref{fig:exp-fit} (b). Again we point that only the
additional restrictions from the neutron channel may allow to discriminate among
different variants.

Important constructive result of our analysis is that so far we have found only
the $T=5/2$ interactions providing $E_r(1/2^-) \gtrsim 2.3$ MeV in $^{9}$He are tolerated by
the data of \cite{Uberseder:2016}. This is consistent with the data of
\cite{Golovkov:2009}, providing $E_r(1/2^-) = 2.0 \pm 0.2$ MeV, but not with
any other $p_{1/2}$ resonance result, see, e.g., \cite{Kalanee:2013} and Refs.\
therein.


\textit{Conclusions.}
%
%
Resonance proton scattering of exotic nuclide on a thick target $^AZ$+$p$ in
inverse kinematics can be used to infer the properties of the low-energy
neutron scattering $^AZ$+$n$ on this nuclide by assuming the isobaric symmetry.
However, for the relatively broad states and the $^AZ$ subsystem with nonzero
isospin (which is always true for exotic dripline nuclei) the effects of the
isospin mixing (due to the $^A(Z+1,\text{IAS})$+$n$ channel) and threshold
effects are large and should be treated dynamically. This introduces in the
situation additional uncertainty: for the $^AZ$+$n$ channel we need only the
nuclear cluster interaction with $T_{\max}$, while for the  $^AZ$+$p$  channel
also the $T_{\max}-1$ interaction should be considered.

The isospin-conserving coupled-channel model is developed to deal with the
problem. The effects  $T_{\max}-1$ are found to be of importance even for
relatively narrow (e.g.\ $\Gamma \sim 0.3$ MeV) \emph{resonant states}; such
treatment is absolutely essential for \emph{broad resonant states} and
\emph{s-wave states}. It was demonstrated that the uncertainty connected with
$T_{\max}-1$ interaction can be overcome only if the outgoing
$^A(Z+1,\text{IAS})$+$n$ channel is studied in parallel with the $^AZ$+$p$
elastic channel. There is an example of such neutron channel studies
\cite{Rogachev:2004}, but these are studies of narrow states and without simultaneous 
consideration of the proton channel.

It is predicted within the ICM that the phase shifts of $T_{\max}=5/2$ and
$T_{\max}-1=3/2$ have very distinct behavior. The commonly expected resonant
phase shift behavior --- with a steep rise and passing $\pi/2$ --- is found common for
$T=3/2$ resonances, not for $T=5/2$. This means that it is likely that
resonances of $^{8}$He+$p$ identified by this standard approach as $T=5/2$ in
\cite{Rogachev:2003a} are actually misidentified. The more recent $^{8}$He+$p$
data of \cite{Uberseder:2016} can be interpreted within ICM in an alternative 
and
more ``orthodox'' way: just weak attraction in the $1/2^+$ channel 
instead of $s_{1/2}$ resonance, single-particle
$1/2^-$ resonance at $E_r = 2.3$ MeV, $5/2^+$ resonance at $E_r = 4.7$ MeV. This
interpretation is consistent with typical theoretical vision of the $^{9}$He spectrum and
especially favors the data of \cite{Golovkov:2009}, which are the only data
positioning the $1/2^-$ state sufficiently high in energy. So, it is demonstrated in this work 
that the results of the resonance proton scattering experiments aiming the studies of the "isobaric partner"
neutron scattering channel may be interpreted in a very different way, 
when isospin conservation is taken properly into account.




\textit{Acknowledgments.}
%
%
We are grateful to Profs.\ E.Yu.~Nikolskii and A.S.~Fomichev for important
discussions. This work was partly supported by the Russian Science Foundation
grant No.\ 22-12-00054.  The research was supported in part in the framework of
scientific program of the Russian National Center for Physics and Mathematics,
topic number 6 ``Nuclear and radiation physics'' (2023--2025 stage).
This work was partly supported by MEYS Project LM2023060.
%
%


\bibliographystyle{apsrev4-2}
\bibliography{isospin-9he-s-3.bbl}
\end{document}